\documentclass[aps,prl,twocolumn,groupedaddress]{revtex4}
\usepackage{graphicx}
\usepackage{amsmath,amsfonts,amssymb}
\usepackage{color}
\usepackage{float}
\begin{document}
\title{Is spacetime absolutely or just most probably Lorentzian?}

\author{Aharon Davidson}
\email{davidson@bgu.ac.il}
%\homepage[}
\author{Ben Yellin}
\email{yellinb@post.bgu.ac.il}
\affiliation{Physics Department, Ben-Gurion University
of the Negev, Beer-Sheva 84105, Israel}

\date{May 1, 2016}

\begin{abstract}
Pre-gauging the cosmological scale factor $a(t)$ does not
introduce unphysical degrees of freedom into the exact
FLRW classical solution.
It seems to lead, however, to a non-dynamical mini superspace.
The missing ingredient, a generalized momentum enjoying
canonical Dirac (rather than Poisson) brackets with the
lapse function $n(t)$, calls for measure scaling which can be
realized by means of a scalar field.
The latter is essential for establishing a geometrical connection
with the 5-dimensional Kaluza-Klein Schwarzschild-deSitter
black hole.
Contrary to the Hartle-Hawking approach,
(i) The $t$-independent wave function $\psi(a)$ is traded for
an explicit $t$-dependent $\psi(n, t)$,
(ii) The classical FLRW configuration does play a major role
in the structure of the 'most classical' cosmological wave packet,
and
(iii) The non-singular Euclid/Lorentz crossovers get quantum
mechanically smeared.

%\medskip
%{\sf{Honorable mention, Gravity Research Foundation (2015)}}
\end{abstract}

%\pacs{}

\maketitle

\noindent \textbf{Mini introduction}\smallskip

Quantum gravity has become the holy grail of theoretical
physics.
The most popular candidates for such an elusive theory
include the canonical Wheeler-DeWitt equation, superstring
theory, loop quantum gravity, non-commutative geometry,
and Regge-type spacetime discretization.
At the bottom line, however, as reflected by the diversity
of the theoretical ideas floating around, quantum gravity
is still at large.
Under such circumstances, exciting only a finite number
of degrees of freedom in superspace \cite{mini} (by
imposing symmetry requirements upon spacetime metrics)
may hopefully allow us to gain some insight \cite{validity}
into quantum gravity, quantum cosmology \cite{QCreview}
in particular.
Hartle-Hawking no-boundary proposal \cite{noboundary}
and the consequent Vilenkin tunneling proposal \cite{tunneling}
constitute such prototype cosmological mini superspace
examples.
It is within the framework of a particular dilaton gravity mini
superspace model (tightly connected with an underlying
Kaluza-Klein black hole configuration), and without challenging
the Hartle-Hawking approach, that we attempt to address a
fundamental quantum gravitational issue.
Namely, is spacetime strictly or just most probably Lorentzian?
The model is presumably over simplified for discussing such
a question, but as a matter of principle, once gravity meets the
quantum world, the possibility that spacetime is just most probably
Lorentzian appears to be quite natural (having in mind that the
complementary probability that spacetime is Euclidean must be
vanishingly small today).

Let us start with the general relativistic action
\begin{equation}
	{\cal S}=-\frac{1}{8\pi}
	\int ({\cal R}+2\Lambda)\sqrt{-g}~d^4 x ~,
	\label{GR}
\end{equation}
$\Lambda$ denoting a cosmological constant.
Associated with the cosmological FLRW line element
\begin{equation}
	ds^2=-n^2(t)dt^2+a^2(t)\left(
	\frac{d r^2}{1-k r^2}+r^2 d\Omega^2
	\right) ~,
\end{equation}
and up to a total time derivative term, is then the well known
mini superspace Lagrangian
\begin{equation}
	{\cal L}_0 (n,a,\dot{a})=\frac{a}{n}\dot{a}^2
	+n a\left(\frac{\Lambda}{3} a^2-k\right)~.
\end{equation}
Treating the scale factor $a(t)$ and the lapse function
$n(t)$ as canonical variables, on an equal footing, leaves
us eventually with a single classical equation of motion
\begin{equation}
	\frac{\dot{a}^2}{n^2}+k=\frac{1}{3}\Lambda a^2 ~,
	\label{FLRW}
\end{equation}
accompanied by an apparent gauge freedom yet to be 
exercised.
Post-gauging the lapse function, that is setting (say) $n(t)=1$,
gives rise to the FLRW equation in its standard differential
form.
The less familiar post-gauging of the scale factor, that is
setting (say) $a(t)=t$ (note that $a(t)$ cannot be a constant
as otherwise $n(t)$ is left undetermined), appears to be
equally permissible even though it drives the FLRW
equation algebraic.

\begin{figure}[]
	\center
	\includegraphics[scale=0.35]{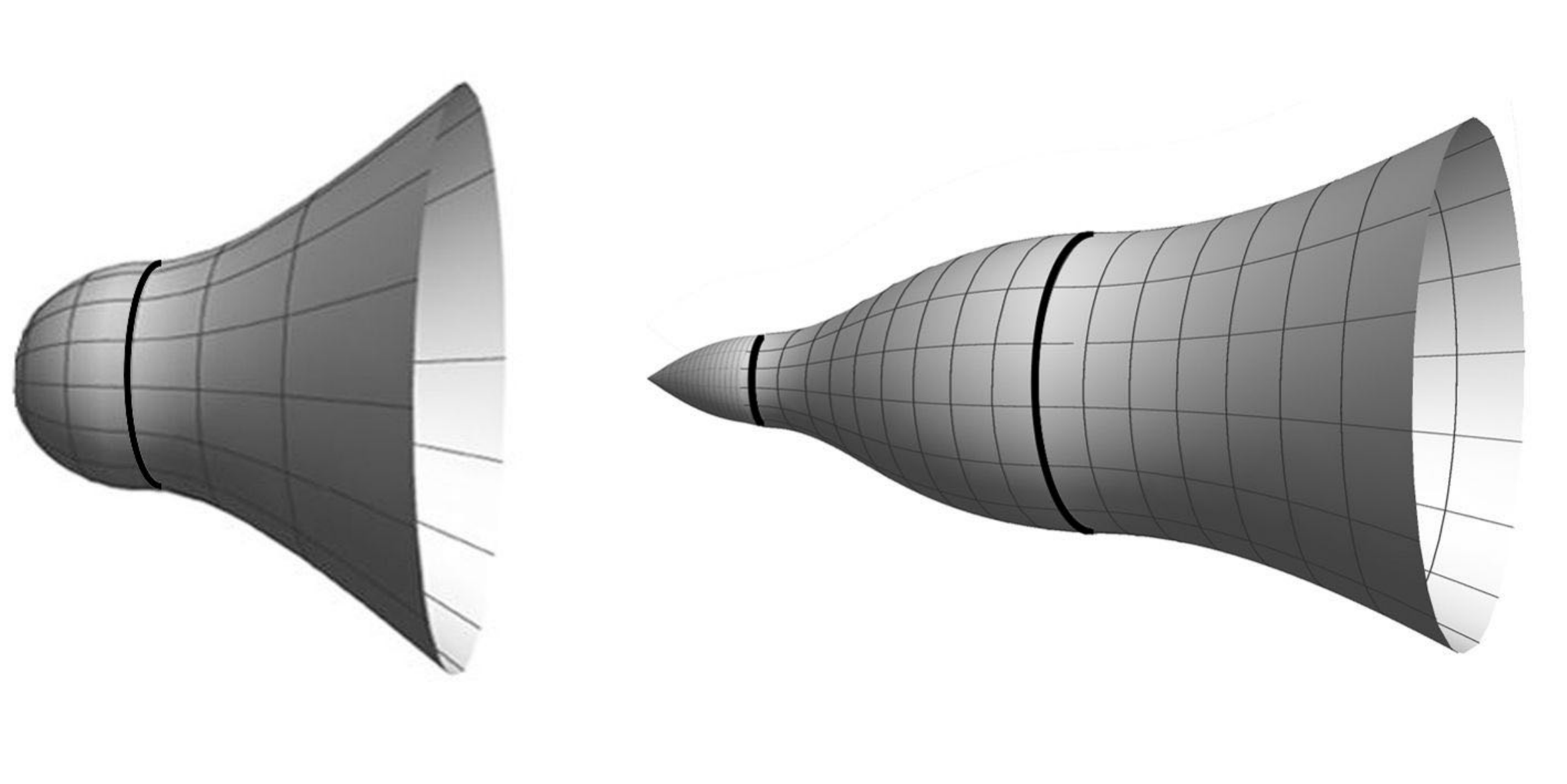}
	\caption{The Hartle-Hawking no-boundary manifold
	(left) depicts a Euclidean sector glued to the Lorentzian
	sector at the so-called 'creation surface'.
	The Vilenkin extension (right) depicts a Euclidean sector
	sandwiched between two (an embryonic and an eternally
	expanding) Lorentzian sectors.}
	\label{Fig1}
\end{figure}

Quantum cosmology, as formulated at the
mini superspace level by means of the constrained
Hamiltonian
\begin{equation}
	{\cal H}(n,a,p_a) =\frac{n}{a}
	\left(\frac{1}{4}p_a^2+a^2 
	(k-\frac{\Lambda}{3}a^2)\right)~,
\end{equation}
is governed by the $t$-independent wave function
$\psi(a)$ obeying the Wheeler-DeWitt (WDW) equation
\begin{equation}
	{\cal H}\psi=0.
\end{equation}
The Hartle-Hawking no-boundary proposal further
calls for $k,\Lambda>0$.
Creation, using the no-boundary proposal language, is then
nothing but a non-singular Euclidean to Lorentzian crossover.
The associated no-boundary manifold is depicted in fig.(\ref{Fig1})
along with the Vilenkin extension \cite{tunneling}.

%% Fig1 %%%
%%%%%%
\smallskip
Two characterisic features are encountered:

\noindent
(i) Ironically, while classical cosmology deals by definition with
the time evolution of the universe, quantum cosmology is apparently
'static'.
What we mean here by 'static' is merely at the technical level.
While classically ${\cal L}(x,\dot{x})$ usually leads to $x(t)$ and
quantum mechanically to $\psi(x,t)$, one would naively expect the
classical $a(t)$ to be traded quantum mechanically for $\psi(a,t)$
rather than by $\psi(a)$.
This indicates that the cosmological wave function does not
have any preferred time direction (and most importantly, should
not be translated into a static universe).
Such an interpretation has initiated a debate regarding both the
notion of time \cite{Rovelli} as well as the arrow of time \cite{Kiefer}
in quantum cosmology.

\noindent
(ii) Quantum mechanically,  for any given scale factor $a$ (not to
confuse with classical $a(t)$), spacetime can be either strictly
Lorentzian or else strictly Euclidean.
The possibility that spacetime is only most probably
Lorentzian, a possibility which cannot be a priori ruled out in
any quantum gravitational model, is absent.

\medskip
\noindent \textbf{Harmful pre-gauging}
\smallskip

In contrast with the post-gauging option, which is realized
at the level of the equations of motion, pre-gauging is carried
out already at the level of the Lagrangian.
The trouble is that had we pre-gauged the lapse function,
pre-fixing for example $n(t)=1$ to start from
\begin{equation}
	{\cal L}_1 (a,\dot{a})=a\dot{a}^2
	+a\left(\frac{1}{3}\Lambda a^2-k\right) ~,
	\label{L1}
\end{equation}
we would have encountered a major problem since
only the 2nd-order Friedmann equation, namely
\begin{equation}
	2\frac{\ddot{a}}{a}+\frac{\dot{a}^2+k}{a^2}
	=\Lambda ~,
\end{equation}
is recovered.
A subsequent integration leads to
\begin{equation}
	\frac{\dot{a}^2+k}{a^2}
	=\frac{1}{3}\Lambda+\frac{E}{a^3}~,
\end{equation}
telling us that it is only when the conserved energy $E$
of the system happens to vanish that one correctly recovers
the missing 1st-order Friedmann equation.
A fake 'dark matter' component, parametrized by $E$, has
entered the game.
In other words, over-gauging has killed the Hamiltonian
constraint (in charge of enforcing $E=0$).
Appreciating this point, the mini Lagrangian eq.(\ref{L1})
cannot be a tenable starting point for a quantum mechanical
cosmological model.

While this drawback is generic, there exists a special pre-gauging
counter example capable of re-producing the exact classical
solution without introducing any fake degree of freedom.
We are talking about pre-gauging the scale factor $a(t)$.
For example, set $a(t)=t$, giving rise to the mini Lagrangian
variant
\begin{equation}
	{\cal L}_2 (n,t)=\frac{t}{n}
	+n t\left(\frac{\Lambda}{3} t^2-k\right) ~.
	\label{L2}
\end{equation}
Counter intuitively, such an unconventional pre-gauging
\cite{essay} does constitute a physically viable option.
Indeed, the reduced Euler-Lagrange equation
$\displaystyle{\frac{\partial {\cal L}_2}{\partial n}=0}$,
albeit an algebraic rather than a differential equation, gives
rise to the exact solution of the full theory, free of any fake
degree of freedom.
To be explicit,
\begin{equation}
	ds^2=-\frac{dt^2}{\frac{1}{3}\Lambda t^2-k}
	+ t^2\left(\frac{d r^2}{1-k r^2}+r^2 d\Omega^2
	\right)~.
\end{equation}
The coefficient of $dt^2$ changes sign from $+\infty$ to
$-\infty$ (crucially without crossing zero).
While the metric is Lorentzian for $t>t_c$, it is strikingly
Euclidean for $t<t_c$.
The pole at
\begin{equation}
	t_c=\sqrt{\frac{3k}{\Lambda}}
\end{equation}
deserves some attention.
Note that all curvature scalars are non-singular everywhere,                                                                                                                                                                                                                                                                                                                                                                                                                                                                                                                                                                                                                                                                                                                                                                                                                                                                                                                                                                                                                                                                                                                                                                                                                                                                                                                                                                                                                                                                                                                                                                                                                                                                                                                                                                                                                
\begin{equation}
	{\cal R}=-4\Lambda ~,~
	{\cal R}^{\mu\nu}{\cal R}_{\mu\nu}
	=4\Lambda^2  ~,~
	{\cal R}^{\mu\nu\lambda\sigma}
	{\cal R}_{\mu\nu\lambda\sigma}
	=\frac{8}{3}\Lambda^2   ~,
	\label{curvature}
\end{equation}
including in particular at $t=0$ and at $t=,t_c$.
This allows for the well known Hartle-Hawking smooth
gluing of the Euclidean and the Lorentzian sub-manifolds
precisely at $t=t_c$ (the manifold is depicted in fig.(\ref{Fig1}).
This point is to be revisited once the lapse function gets
slightly modified.

To see things from a different perspective, one may follow
Dirac formalism \cite{Dirac} for dealing with constrained
systems.
As long as ${\cal L}_2 (n,t)$ is not a linear function
of $n$, the classical equations of motion residing from
the corresponding total Hamiltonian serve
as consistency checks for the two (a primary and a secondary)
second class constraints
\begin{equation}
	p_n\approx 0~, \quad
	\left\{p_n, {\cal H}_2\right\}_{_P}
	=\frac{\partial {\cal L}_2}{\partial n}\approx 0~.
\end{equation}
The catch is, however, that starting from ordinary 
Poisson brackets
\begin{equation}
	\left\{n,p_n \right\}_{_P}=1 ~,
\end{equation}
one ends up with vanishing Dirac brackets
\begin{equation}
	\left\{n,p_n \right\}_{_D}=0 ~.
	\label{vanishD}
\end{equation}
The lesson is clear: Re-producing the exact classical solution
is just a necessary condition any mini superspace model must
obey.
The trouble is that it is not always a sufficient condition as well.
As indicated by the vanishing Dirac brackets, the mini superspace
model based on the Lagrangian eq.(\ref{L2}) is non-dynamical.
While classically this is tolerable, the standard quantization
procedure, based on replacing the canonical Dirac brackets by
quantum mechanical commutation relations, is blocked.
There are cases, however, see the forthcoming dilaton model,
where tenable pre-gauging turns harmless, at least in the sense
that it leads to a dynamical mini superspace.

\medskip
\noindent \textbf{Tenable pre-gauging}\smallskip

In search for the missing ingredient, namely a generalized
momentum $\tilde{p}_n$, Dirac (rather than Poisson) conjugate
to the lapse function $n$, we have come across the modified
gravity action
\begin{equation}
	{\cal S}=-\frac{1}{8\pi}
	\int ({\cal R}+4\Lambda)\phi
	\sqrt{-g}~d^4 x ~,
	\label{scaleaction}
\end{equation}
involving a conformally coupled scalar field $\phi$
(the factor 4 is to be apostriori justified by eq.\ref{rhop}).
In a somewhat different language, the above prescribes
a scale modification $\sqrt{-g}\rightarrow \phi \sqrt{-g}$
of the general relativistic measure.
The Kaluza-Klein perspective will be discussed in a forthcoming
section.

A few remarks are in order:

\noindent (i) One may think of $\phi$ in terms of
$\phi=\varphi^{\dagger}\varphi $.
In fact, this is a convenient way to explicitly break the
classical symmetry $\phi \rightarrow -\phi$ of the field
equations, thereby getting rid of the potentially problematic
$\phi<0$ anti-gravity \cite{GaG} regime.

\noindent (ii) It should be emphasized that also the current
scheme, like the Hartle-Hawking scheme, also has no pretensions
to go beyond the scope of a early universe model.

\noindent (iii) Reflecting its linearity in the dilaton field,  the
action eq.(\ref{scaleaction}) has no $f({\cal R})$ analogous theory
\cite{fR}.

\noindent (iv) The justification for our modified action comes
from the Kaluza-Klein theory.
Eq.(\ref{scaleaction}) is nothing but the 4-dim reduction of
a 5-dim Einstein-Hilbert action supplemented by a cosmological
constant.
We later show how the associated 4-dim cosmological evolution,
differing from the standard FLRW evolution, is intimately connected
with the de-Sitter Kaluza-Klein black hole.

The special dilaton field $\phi(t)$ is accompanied by a
linear scalar potential $V(\phi)=4\Lambda\phi$, and is
stripped from any Brans-Dicke kinetic term.
It is nevertheless a dynamical scalar field as can be verified
by extracting its hidden Klein Gordon equation
\begin{equation}
	\frac{1}{n a^3}\frac{d}{dt}\left( \frac{a^3}{n}\frac{d
	\phi}{dt} \right)+V^\prime_{eff}(\phi) =0~,
	\label{KG}
\end{equation}
from the two independent classical equations of motion
involved.
To be more specific, the time evolution of the dilaton
is governed in our case by the effective upside down harmonic
potential
\begin{equation}
	V_{eff}(\phi)=\frac{1}{3}\int
	(\phi V^\prime-2V)d\phi=-\frac{2}{3}
	\Lambda\phi^2+const ~,
	\label{Veff}
\end{equation}
rather than by the original linear potential itself.
Note that $\phi=const$ is not a solution.
That is to say that the model, at least in its current form, lacks
a general relativistic limit.
Incorporating a proper general relativistic limit would only require
the introduction of a more elaborated effective potential,
say $V(\phi)=4\Lambda\phi+\lambda\phi^2+\gamma\phi^3$,
with $16\pi\langle\phi\rangle$ then serving as the reciprocal Newton
constant.
Such a general relativistic limit is known to be characterized
by matter dominated (in average) Hubble constant oscillations
\cite{averageH}.
The general relativistic late universe is, however, beyond the
scope of our early universe dilaton model (and Hartle-Hawking
model as well).

The cosmological evolution, characterized by a
constant Ricci scalar ${\cal R}=-4\Lambda$ in the Jordan
frame, is then translated  \cite{Tomer} into the equation
of state
\begin{equation}
	\rho-3P=4\Lambda ~,
\end{equation}
 which subject to energy/momentum conservation implies
\begin{equation}
	\rho=\Lambda +\frac{\xi}{a^4} ~, \quad
	P=-\Lambda +\frac{\xi}{3a^4}~.
	\label{rhop}
\end{equation}
The constant of integration $\xi$ signals the presence of
a radiation-like term, the fingerprint of measure scaling.
Note that so far, in the literature, radiation has only been
introduced ad-hoc \cite{Vilenkin} into the mini superspace
model, inducing (for $\xi>0$) an embryonic universe epoch.
Eventually, either recovering from a radiation dominated
Big-Bang or else bouncing from a potential hill, the universe
glides asymptotically into the de-Sitter phase.

Switching on now the scale factor pre-gauging $a(t)=t$, the
mini superspace Lagrangian ${\cal L}_2(n,t)$ gets slightly
yet significantly modified to read
\begin{equation}
	{\cal L}(n,\phi,\dot{\phi},t)
	=\frac{t}{n}(\phi+t\dot{\phi})
	+nt\left(\frac{2}{3}\Lambda t^2-k\right)\phi~.
	\label{Lt}
\end{equation}
One may verify that the equations of motion derived from
this mini Lagrangian give rise to the exact classical solution
of the full dilaton gravity theory, namely
\begin{equation}
	n^{-2}(t)=\frac{1}{3}
	\Lambda t^2-k+\frac{\xi}{t^2} ~,
	\quad n^2 (t)\phi^2 (t)=const ~,
	\label{nphi}
\end{equation}
without introducing any fake degree of freedom.
If, in addition to the Hartle-Hawking requirements $\Lambda,k>0$,
we further require $0<\xi<\frac{3k^2}{4\Lambda}$, thereby allowing
$n^{-2}(t)=0$ to admit two positive roots $0<t_-<t_+$.
The emerging Euclidean sector  associated with  $t_-<t<t_+$
serves as a classical barrier disconnecting a Lorentzian Embryonic
universe ($0<t<t_-$) from the expanding $\Lambda$-dominated
Lorentzian universe ($t_+<t<\infty$).

The various curvature scalars eq.(\ref{curvature}) are respectively
modified to  
\begin{eqnarray}
	&&{\cal R}=-4\Lambda ~,~
	{\cal R}^{\mu\nu}{\cal R}_{\mu\nu}
	=4\Lambda^2+\frac{12\xi^2}{t^8}  ~,~
	\nonumber \\
	&&\quad{\cal R}^{\mu\nu\lambda\sigma}
	{\cal R}_{\mu\nu\lambda\sigma}
	=\frac{8}{3}\Lambda^2+\frac{24\xi^2}{t^8}   ~.
\end{eqnarray}
For $\xi \neq 0$, at $t\rightarrow 0$, as expected, the singular
Big-Bang event makes its unavoidable appearance.
On the other hand, however, the Euclidean to Lorentzian
transitions associated with the lapse function pole behavior at
\begin{equation}
	t^2_{\pm}=\frac{3k}{2\Lambda}\left(
	1 \pm \sqrt{1-\frac{4\Lambda\xi}{3k}}\right)~,
\end{equation}
stay curvature non-singular. This is a crucial point.
In the Hartle-Hawking case, $n^2(t)<0$ marks a classically
forbidden territory.
The more so in the present case, where $n^2(t)<0$ is further
accompanied by $\phi^2(t)<0$.
In particular, one faces $\phi^2(t_{\pm})=0$ at creation.
We will re-visit this intriguing $n(t) \leftrightarrow \psi(t)$
sign interplay once the geometrical connection with Kaluza-Klein
black hole is established.

The point is that the Lagrangian eq.(\ref{Lt}), contrary to the
problematic Lagrangian eq.(\ref{L2}), does lead to a dynamical
mini superspace (with pre-gauging involved, perhaps a better
name is micro-superspace).
We proceed now to show how measure scaling combined
with scale factor pre-gauging opens the blocked trail to quantum
cosmology.

\medskip
\noindent \textbf{From $\psi(a)$ to $\psi(n,t)$}

Within the framework of the Hamiltonian formalism, the
fact that one cannot express the 'velocities' $\dot{n},\dot{\phi}$
in terms of the associated momenta
$\displaystyle{p_n=\frac{\partial {\cal L}}{\partial \dot{n}}}$
and $\displaystyle{p_{\phi}=
\frac{\partial {\cal L}}{\partial\dot{\phi}}}$ results
in two primary second class constraints
\begin{eqnarray}
	&\displaystyle{\chi_n \equiv p_n\approx 0~,
	\quad \chi_{\phi} \equiv
	p_{\phi}-\frac{t^2}{n}\approx 0~,}&\\
	&\displaystyle{\{\chi_n, \chi_{\phi}\}_{_P}
	=\frac{t^2} {n^2}~.}&
\end{eqnarray}
Following Dirac prescription, one can keep using the
Poisson brackets provided the naive Hamiltonian ${\cal H}$
is traded for the total Hamiltonian
\begin{equation}
	{\cal H}_T={\cal H}+u_n \chi_n+u_{\phi}\chi_{\phi}~,
\end{equation}
with the two coefficients $u_{1,2}$ fixed on consistency grounds by 
the requirements
\begin{equation}
	\{\chi_{1,2},{\cal H}_T \}_{_P}+
	\frac{\partial \chi_{1,2}}{\partial t}=0
\end{equation}
Alternatively, one may stick to the naive Hamiltonian
\begin{equation}
	{\cal H}=-\phi t\left[
	\frac{1}{n}+n\left(\frac{2}{3}\Lambda t^2-k\right)
	\right] ~,
\end{equation}
but pay the price of replacing the Poisson brackets by
the Dirac brackets.
Obviously, only the latter option is relevant for the
quantization procedure which is carried out by means
of $\{A,B\}_{_D}=1 \rightarrow [A,B]=i \hbar$.
With this in mind, we find it now convenient to define
a new set of canonical variables
\begin{equation}
	x(t)\equiv \frac{t^2}{n^2(t)}~, \quad
	y(t)\equiv \phi^2(t)t^2~,
\end{equation}
which are introduced of course already at the level of the
mini superspace Lagrangian.
Such a special  $(x,y)$-representation, for which the naive
Hamiltonian acquires the compact, notably momentum-free,
form
\begin{equation}
	{\cal H}(x,y,t)=-t \left(\frac{2}{3}
	\Lambda t^2-k\right)\sqrt{\frac{y}{x}} ~,
\end{equation}
has been especially designed to remove any explicit time
dependence from the various Dirac brackets involved.
The crucial observation now is that
\begin{equation}
	\{x,y\}_{_P}=0
	\quad \Longrightarrow\quad
	\{x,y\}_{_D}\neq 0 ~,
\end{equation}
a fact which has far reaching consequences on the
quantization procedure.
The WDW wave function cannot depend on $x$ and $y$
simultaneously.
To be more specific,
\begin{equation}
	\{x,y\}_{_D}=-4\sqrt{xy} 
	\quad \Longrightarrow\quad
	\left\{x,-\frac{1}{2}\sqrt{\frac{y}{x}}\right\}_{_D}=1~.
\end{equation}
Measure scaling has eventually given birth to a generalized
momentum Dirac conjugate to the lapse function.
In turn, elevated to the level of a quantum mechanical operator,
$\sqrt{y/x}$ can be faithfully represented by
\begin{equation}
	\sqrt{\frac{y}{x}}=2i\hbar \frac{\partial}{\partial x}~.
\end{equation}

Altogether, up to some c-number (reflecting ambiguities
arising from operator ordering), the Wheeler-DeWitt
Schrodinger equation acquires
the atypical linear form
\begin{equation}
	-2i t\left(\frac{2}{3}\Lambda t^2-k\right)
	\frac{\partial \psi}{\partial x}
	=i \frac{\partial \psi}{\partial t}~,
	\label{WDWS}
\end{equation}
devoid of $\hbar$.
Its most general solution is given by
\begin{equation}
	\psi(x,t)=\psi(x-\frac{1}{3}\Lambda t^4+k t^2)~.
\end{equation}
Not only is the WDW wave function $t$-dependent,
but it furthermore maintains a constant value, namely $\psi(\xi)$,
along the classical path.

One particular solution deserves further attention.
We refer of course to the 'most classical' configuration, namely
a Gaussian wave packet \cite{wavepacket} characterized by
the minimal uncertainty relation, explicitly given by
\begin{equation}
	\boxed {\psi(x,t)=\frac{e^{\displaystyle{-
	\frac{\left(x-x_{cl}(t)\right)^2}
	{4\sigma^2}+\frac{i\eta}{\sqrt{2}}(x-x_{cl}(t))}}}
	{(2\pi)^{\frac{1}{4}}\sqrt{\sigma}}}
	\label{wavepacket}
\end{equation}
factorized by an optional linear phase.
It is by no means trivial, yet consistent with the correspondence
principle (recall the underlying 'most classical' wave packet), that
the most probable as well as the average configuration, is nothing
but the classical FLRW solution
\begin{equation}
	\langle x \rangle=x_{cl}(t)=
	\frac{\Lambda}{3} t^4-k t^2+\xi
	~,~
	\langle p_x\rangle
	=\langle -i\hbar\frac{\partial}{\partial x}\rangle
	=\hbar\eta~.
\end{equation}
The arbitrary constant $\eta$ matches the classical
value of the $n(t) \phi(t)$ product.
Note that the uncertainties are time independent
\begin{equation}
	\Delta x=\sigma~, ~\Delta p_x=\frac{\hbar}{2\sigma} ~.
\end{equation}

\begin{figure}[]
	\center
	\includegraphics[scale=0.55]{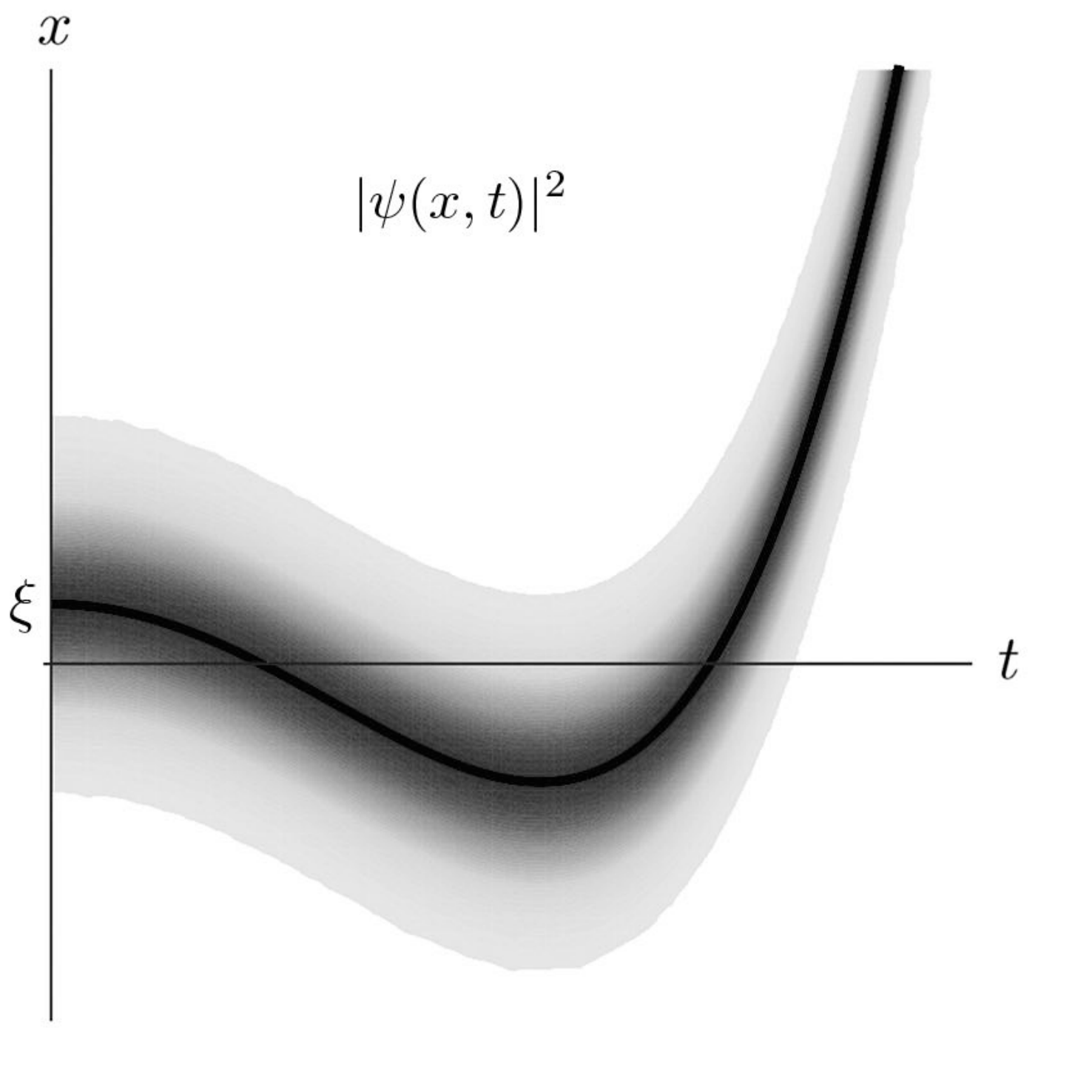}
	\caption{Cosmological wave packet probability density:
	For any time $t$, unlike in the Hartle-Hawking scheme,
	there is a probability that spacetime is Lorentzian ($x>0$)
	and a complementary probability that it is Euclidean
	($x<0$).
	The sharp Euclidean to Lorentzian regime crossover
	($\equiv$ creation) has been quantum mechanically
	smeared.}
	\label{Fig2}
\end{figure}

The cosmological wave packet probability density $|\psi(x,t)|^2$
is depicted in Fig.(\ref{Fig2}).
Roughly speaking, we recognize the mainly Euclidean region
sandwiched between an optional (for $\xi>0$) mainly Lorentzian
Vilenkin-type embryonic era (on the lhs) and the asymptotically classical
mainly Lorentzian region (on the rhs).
As a consistency check one may verify that the probability
$P(t)$ that spacetime is Lorentzian approaches unity as $t\rightarrow\infty$.
To be specific,
\begin{equation}
	\boxed {P(t)=\int_0^{\infty} |\psi(x,t)|^2 dx
	\simeq 1-\frac{3\sigma}{\sqrt{2\pi}\Lambda t^4}
	e^{\displaystyle{-\frac{\Lambda^2 t^8}{18\sigma^2}}}}
	\label{P}
\end{equation}
At this point, noticing that eq.(\ref{WDWS}) is $\hbar$-independent,
one may wonder where is the Planck scale hiding?
On dimensional grounds, it can only enter the game via the spread
parameter $\sigma$.
In fact, as it has been argued, see ref.\cite{area}, based on
micro black hole thermodynamics arguments (the behavior of
the entropy and the Helmholtz free energy as functions of the
Hawking temperature), one may expect
$\sigma \simeq \ell_{Planck}\sim \sqrt{\hbar}$.
This would assure us that a Euclide $\leftrightarrow$ Lorentz
transition is highly improbable at large $t$.
At short $t$, on the other hand, the quantum mechanically
smeared Lorentz $\leftrightarrow$ Euclid interplay takes over.
Even the naive question 'when has creation occurred?' can
only be answered statistically.
This may be connected to the general idea of the multiverse.
The no-boundary proposal limit, if exists, is still hidden at this stage.

%% Fig2 %%%
%%%%%%

\medskip
\noindent \textbf{The Kaluza-Klein black hole connection}

Importing the classical solution
eq.(\ref{nphi}) previously derived into a 5-dimensional world
(with a 3-dimensional maximally symmetric subspace
imposed), we face the line element
\begin{equation}
	ds^2_5=-n^2(t)dt^2
	+t^2\left(\frac{d\rho^2}{1-k \rho^2} 
	+\rho^2 d\Omega^2\right)
	+\phi^2(t)dx^2_5~.
	\label{KK}
\end{equation}
For the sake of definiteness, let us momentarily assume
$\Lambda,k>0$ to keep track with the Hartle-Hawking
scheme, and thereby encounter the familiar
$x_5$-independent (a la Kaluza-Klein by virtue of Birkhoff
theorem) general relativistic Schwarzschild-deSitter black
hole metric in disguise.
The standard form is met by simply replacing the notations
according to
\begin{equation}
	t \rightarrow r,~x_5 \rightarrow t~,
\end{equation}
accompanied by the (mass)$^2$ identification $\xi=m^2$.
Altogether, this establishes an intriguing analogy between
the 4-dimensional $t$-dependent quantum cosmology and
the 5-dimensional $\rho$-dependent black hole.
In particular, the Euclidean $\leftrightarrow$ Lorentzian
transition at $t=t_{+}$, referred to as the creation
surface in the 4-dim language, is recognized as the outer
horizon of the Schwarzschild-de-Sitter Kaluza-Klein black hole.
This is of course not a coincidence.
By substituting the 5-dim metric ansatz eq.(\ref{KK}) into
the Kaluza-Klein action
\begin{equation}
	{\cal S}_5=-\frac{1}{4\pi}\int
	\left({\cal R}_5+4\Lambda \right)
	\sqrt{-g_5}~d^5x ~,
\end{equation}
and integrating out the $S_3$ sub-manifold as well as the
$x_5$-circle, one encounters (up to a total derivative) the
one and the same mini superspace Lagrangian eq.(\ref{Lt}).

With this in mind we note that the quantum mechanical
4-dim Schwarzschild black hole has already been discussed
in the midi-superspace model \cite{Kuchar}, in the Dirac
style pre-gauging formalism \cite{area}, and in the technical
context of a generalized uncertainty principle \cite{Casadio}.
The 'most classical' non-singular circumferential radius
$r$-dependent Schwarzschild black hole wave packet is
given by
\begin{equation}
	\psi_{BH}(x,r)=\frac{1}{(2\pi)^{\frac{1}{4}}\sqrt{\sigma}}
	e^{-\displaystyle{
	\frac{\left(x-2r+4m\right)^2}
	{4\sigma^2}}}~.
	\label{psjBH}
\end{equation}
A linear phase is optional.
Up to an $r \leftrightarrow t$ role interchange,
eq.(\ref{psjBH}) highly resembles eq.(\ref{wavepacket}).
In particular, the variable $x$ which has played the role
of $-t^2/g_{tt}$ in the our cosmological scheme, stands
here for $2r/g_{rr}$.
The dictionary continuous as follows.
The analog of the mass parameter $m$ is clearly the
amount $\xi$ of cosmological radiation.
Likewise, the analog of the classical black hole event
horizon (at $r=2m$) is the so-called creation surface (at
$t=t_c$) on which $n^{-2}(t_c)$ happens to vanish.
Carrying the analogy one step further, the cosmological
(black hole) wave packet probability density
$\psi^{\dagger}\psi$ can be effectively translated into
a statistical mechanics radiation (mass) spectrum.

\medskip
\noindent \textbf{Summary table}\smallskip

To summarize, here is a compact table emphasizing
the main differences between the standard Hartle-Hawking mini
superspace approach and our measure scaling dilatonic
(Kaluza-Klein originated) variant:
\begin{equation*}
	\begin{array}{c|c|c}
	& \text{HH model} & \text{This model}\\\hline 
	\text{Wave function} & t\text{--independent} &
	t\text{--dependent}  \\\hline 
	\text{Classical solution} & \text{QM irrelevant} &
	 \text{QM relevant}  \\\hline 
	\text{Black Hole analogy} & \text{No} &
	\text{Yes (5-dim)} \\\hline 
	\text{Euclid/Lorentz crossover}  & \text{Sharp}  &
	\text{Smeared}
	\end{array}
\end{equation*}
A mini superspace model can never replace the full theory.
This is true for the Hartle-Hawking mini superspace model,
and holds of course for the present micro superspace model
as well.
It can at most give us a clue what to expect when quantum
gravity will eventually make its appearance.
And once gravity meets the quantum world (going beyond
the level of various mini superspace models), the idea that
spacetime is just most probably Lorentzian seems quite natural
and in some respect even unavoidable. 
As far as we can tell, such an idea has never been put forwards.
While the complementary probability that spacetime is Euclidean
must be vanishingly small today, it could have had interesting
consequences at the very early universe owing to the quantum
mechanical smearing of the Euclid/Lorentz crossover.

\bigskip
\acknowledgments
{We cordially thank BGU president Prof. Rivka Carmi for
her kind support. A valuable discussion with Prof. Alexander
Vilenkin is very much appreciated.}

\end{document}